\documentclass[twocolumn,english]{revtex4}
\usepackage[T1]{fontenc}
\usepackage[latin9]{inputenc}
\usepackage{amsmath}
\usepackage{graphicx}
\usepackage{amssymb}

\makeatletter
\providecommand{\citet}[1]{\cite{#1}{}}

\usepackage{babel}
\makeatother

\begin{document}

\title{Dicke model: entanglement as a finite size effect }

\author{Oleksandr Tsyplyatyev }

\affiliation{Department of Physics, University of Basel, Klingelbergstrasse 82,
CH-4056 Basel, Switzerland }

\author{Daniel Loss}

\affiliation{Department of Physics, University of Basel, Klingelbergstrasse 82,
CH-4056 Basel, Switzerland }

\begin{abstract}
We analyze the Dicke model at zero temperature by matrix
diagonalization  to determine the entanglement in the ground state.
 In the infinite system limit the mean field approximation
predicts a quantum phase transition from a non-interacting state
to a Bose-Einstein condensate at a threshold coupling. We show
that in a finite system the spin part of the ground state
is a bipartite entangled state, which can be tested by probing
two parts of the spin system separately, but only in a narrow regime around
the threshold coupling. Around the resonance, the 
size of this regime is  inversely proportional to the number of spins
and shrinks down to zero for infinite systems. This spin entanglement
is a non-perturbative effect and is also missed by the mean-field approximation. 
\end{abstract}

\date{\today}

\maketitle
Coherent interaction between electromagnetic photon fields and matter attracted
interest a long time ago \cite{Gardiner} with renewed
attention gained in the last decade due to significant developments
in the experimental techniques  in various areas of physics.
Achievement of Bose Einstein condensation of cold atomic gases in
electromagnetic traps enabled the coherent coupling of
hyperfine states of $10^{5}$ atoms to a single photon mode of
an optical resonator \cite{esslinger}. Advances in the
semiconductor technology allowed to obtain optical microcavities
where electron-hole excitations inside
the semiconductor quantum well are strongly coupled to an eigenmode
of the optical resonator \cite{kasprzak}. Strong coupling
of a single mode of a transmission line resonator to a Cooper pair
box \cite{wallraff} and a quantized mode of an optical crystal cavity
to several semiconductor quantum dots \cite{Badolato1}
have been demonstrated as a possible way to a quantum computing device \cite{AwschalomQED}. 

Theoretical understanding of all these systems is based on a model
proposed by Dicke \cite{Dicke} which describes $N$ spins 1/2 (identical
two-level systems) with splitting energy $2\epsilon$ coupled to a
single mode of electromagnetic field $\omega$. It was shown that
this model is exactly diagaonalisable \cite{TavisCummings}. At zero
temperature it undergoes a quantum phase transition from a non interacting
state with unpopulated bosonic mode to a condensed state with with a highly populated  bosonic mode  \cite{scharf} if coupling between the
boson and a single spin $g$ is greater than a threshold value.
In the thermodynamic limit a phase transition occurs in the region
of strong coupling if temperature is less than a critical temperature
which can be described by a Bogolyubov Hamiltonian similarly to the
pairing model of superconductivity \cite{HeppLieb}. Recently, a variational
wave function approach to the generalised Dicke model was used \cite{eastham}
to describe Bose Einstein condensation of exciton-polaritons in a
semiconductor optical cavity.

In this paper we analyse the Dicke model at zero temperature for a
finite $N$ using matrix diagonalisation methods. We find that for the particular
coupling strength, \begin{equation}
\omega\epsilon<g^{2}\lesssim\omega\epsilon\left(1+\frac{1}{N}\right),\label{eq:g_1boson}\end{equation}
the ground state of the spin subsystem is a bipartite
entangled state and it is not entangled outside of this region. The lower bound of the inequality is onset of the
quantum transition described by the mean field theory \cite{MeanField}.
The upper bound is the condition to have only singly populated bosonic mode 
in the ground state. The approximated value is the result of
$1/N$ expansion around the resonance. 

In the thermodynamic limit the ground state  projected
onto the subspace of N spins is not entangled as it is a product state. 
For a weak coupling below the quantum transition
the ground state is a product of all unexcited single spin states.
For a strong coupling above the transition threshold the ground state
is also a product state of all single spin states \cite{eastham} as a result of the mean-field approximation.
In a finite system the mean-field approximation is not applicable
in a small region above the transition threshold where the expectation
value of the boson is of the order of 1 and its fluctuations are also
of the order of 1. The ground state in this region is
a superposition of the unexcited state of all spins and a spin state
with only one single spin-flip excitation, the $N$ spin W-state, $\left|W\right\rangle=\left( \left| \uparrow\downarrow\downarrow\dots\right\rangle+\left| \downarrow\uparrow\downarrow\dots\right\rangle+\left| \downarrow\downarrow\uparrow\dots\right\rangle+\dots \right)/\sqrt{N}$. The W-state can also be interpreted as 'magnon state' at vanishing wave vector \cite{Pratt}.

Furthermore, the W-state can be considered as a bipartite entangled state in the following sense. Dividing all spins into
two groups \cite{spin_division} the W-state is a Bell state
in the subspace restricted by only one spin-flip excitation above
the unexcited states of each group of the spins, see Eq.~(\ref{eq:m1m2}).
In the course of a bipartite measurement if the first group is found
in the excited state then the second group is projected onto the unexcited
state and if the first group is found in the unexcited state then
the second group is projected onto the excited state.

We diagonalize  the Dicke model  for $N$ spins 1/2 coupled to a single bosonic mode 
\begin{equation}
H=\omega b^{\dagger}b+\epsilon\sum_{j}S_{j}^{z}+\frac{g}{\sqrt{N}}\sum_{j}
\left( S_{j}^{+}b+S_{j}^{-}b^{\dagger} \right), \label{eq:Dicke_Model}
\end{equation}
where the sum runs over $N$ spin-1/2 operators 
${\bf S}_j$ that obey the commutation relations 
$ [ S_i^\alpha , S_j^\beta ] = \epsilon_{\alpha \beta \gamma} \delta_{ij} S_i^\gamma$,
and $b$ $(b^\dagger)$ is standard bosonic annihilation (creation) operator.

The Dicke model possesses the following conserved quantities. One is the number of excitations
of the coupled spin-boson system,
\begin{equation}
L=n+J_{z},\label{eq:cooperativity}\end{equation}
 expressed in terms
of the z-component of total spin operator $J_{\alpha}=\sum_{j}S_{j}^{\alpha},  \alpha=x,y,z$,  and 
the occupation number operator $n=b^{\dagger}b$ of the boson mode. Note that $J_{\alpha}$ and $n$ are not conserved separately.
The eigenvalues of $L$ are the so-called cooperation numbers $c$, given by the sum of expectation values 
of $n$ and $J_{z}$. 
A second conserved quantity is the total spin, 
$J^{2}=J_{z}^{2}+\left(J_{+}J_{-}+J_{-}J_{+}\right)/2$ with eigenvalues $j\left(j+1\right)$.

We represent  the Dicke Hamiltonian (\ref{eq:Dicke_Model}) in a basis
where $J^{2}$ and  $L$  are block-diagonal.
 Within a block of given $c$ and $j$ the remaining degrees of freedom can
be labeled by the  eigenvalues $m$ of $J_{z}$. In the representation $\left|c,j,m\right\rangle $
each block has a tridiagonal form. The diagonal matrix elements
represent the energies of states containing $c-m$ bosons and $m$ excited spins, \begin{equation}
\left\langle m\middle|H\middle|m\right\rangle =\omega\left(c-m\right)+\epsilon m.\label{eq:diagonal_ME}\end{equation}
The first above- and below-diagonal matrix elements are transition
amplitudes connecting all pairs of states which differ by just one flipped spin,
\begin{equation}
\left\langle m\middle|H\middle|m+1\right\rangle =\frac{g}{\sqrt{N}}\sqrt{\left(c-m\right)\left(j\left(j+1\right)-m\left(m+1\right)\right)}.\label{eq:offdiagonal_ME}\end{equation}
The size of each block is limited by the fact that $-N/2\leq m \leq N/2$ for $j=N/2$. However,the upper bound on $m$ is further constrained by the cooperation 
number $c$ which has the lower bound $-N/2$ for a block with no
bosons present in the state.

The ground state of Eq.~(\ref{eq:Dicke_Model}) is the lowest energy
state of all blocks Eqs.~(\ref{eq:diagonal_ME},\ref{eq:offdiagonal_ME}).
For different values of the parameters $\omega$, $\epsilon$,
and $g$ the ground state can have  different $c's$ and  $j's$ (see below).

\begin{figure}
\centering\includegraphics[clip,scale=0.45]{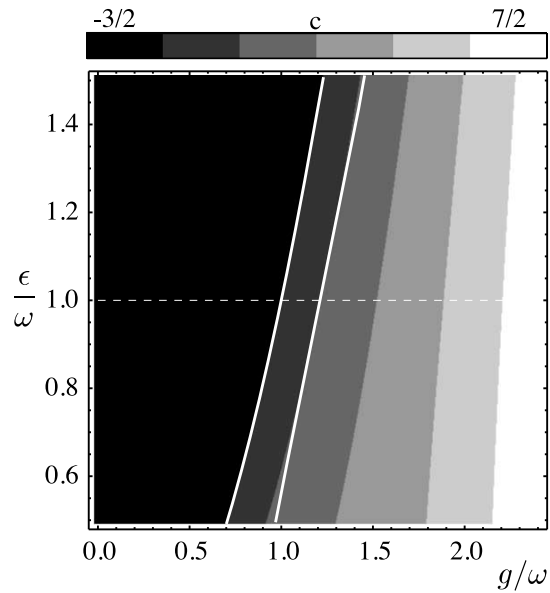}

\caption{\label{fig:gs_cooperativity} Cooperation number $c$ of the ground
state obtained by matrix diagonalization of the Dicke model for $N=3$. The first 
solid line from the left indicates the  quantum
transition to the strongly correlated non-perturbative regime,  and the second solid line is determined by Eq.~(\ref{eq:E2_boundary}).
The dashed line corresponds to the resonance $\epsilon=\omega$.}

\end{figure}

We consider now the case with the bosonic ($\omega$) and spin excitation energies ($\epsilon$) close to each other.
For the uncoupled case, $g=0$, the ground state contains no bosons and no excited spins (i.e. all spins are, say, down).
From Eqs.~(\ref{eq:cooperativity}) and (\ref{eq:diagonal_ME}) we see that
the cooperation number of this state is then $c=-N/2$ with corresponding ground state energy $E_{0}=-N\epsilon/2.$
For finite but still small coupling,
$g\ll1$, we can use a perturbative approach to remove the boson mode 
via a Schrieffer-Wolff transformation to obtain
an effective spin-Hamiltonian  \cite{mircea},
\begin{multline}
H=\omega b^{\dagger}b+\left(\epsilon+\frac{2g^{2}b^{\dagger}b}{N\left(\epsilon-\omega\right)}\right)\sum_{j}S_{j}^{z}\\
+\frac{g^{2}}{N\left(\epsilon-\omega\right)}\sum_{i,j}S_{i}^{+}S_{j}^{-},\label{eq:H_SW}\end{multline}
where the spin-boson mixing is eliminated up to the second order
in $g$ \cite{SW_details}, introducing an effective XY coupling between the spins
within a subband of given boson occupation number $n$ (with $n$ being conserved under the effective Hamiltonian
Eq.~(\ref{eq:H_SW})). Such a perturbative approach gives a qualitatively
correct description of the energies and wave functions
until $g$ crosses a threshold value $g_c$ where  a  transition 
to a strongly correlated non-perturbative regime takes place (see below).
The cooperation number $c$ of the ground state for the Dicke Hamiltonian Eq.~(\ref{eq:Dicke_Model})
is plotted in Fig.\ref{fig:gs_cooperativity} for the solution for $N=3$. There are several regimes: for small g the ground state of the system
is defined by a regime (black) with $c=-N/2$ and  $\langle n \rangle=0$. Then, with increasing coupling a quantum phase transition at $g_c$ takes place to
a new regime with  $c>-N/2$ where $\langle n \rangle >0$, and where  subbands with sharp bosonic occupation numbers  no longer exist.

Increasing $g$ further above the transition threshold a sequence of states
with $c=1-N/2$,  $c=2-N/2$, and so on becomes subsequently the
ground state due to the interaction energy. The ground state with
$c$ greater but close  to $-N/2$ can not already be approximated with
help from the perturbation theory as coupling is too strong and can
not yet be approximated using the mean field approach as the fluctuations
of the order parameter are too large compared to its mean. Here we
diagonalize the Hamiltonian Eq.~(\ref{eq:Dicke_Model}) using
the matrices Eqs.~(\ref{eq:diagonal_ME},\ref{eq:offdiagonal_ME}).
Sizes of the matrices in this regime are limited by the cooperation number $c$. In the presentation $\left|c,j,m\right\rangle $ the
matrices are not larger then $2 \times 2$ and $3 \times 3$ for $c=1-N/2$
and $c=2-N/2$ respectively. The lowest eigenenergies of these matrices
are 

\begin{equation}
E_{1}=-\frac{\left(N-1\right)\epsilon}{2}+\frac{\omega}{2}-\sqrt{g^{2}+\left(\frac{\omega-\epsilon}{2}\right)^{2}}\label{eq:E1}\end{equation}
for $c=1-N/2$ and \begin{equation}
E_{2}=-\frac{\left(N-2\right)\epsilon}{2}+\omega-2g\sqrt{1-\frac{1}{2N}}+\frac{\omega-\epsilon}{2\left(2N-1\right)}.\label{eq:E2}\end{equation}
for $c=2-N/2$. The last expression is a result of expansion at the
resonance in powers of $\omega-\epsilon$. Both $E_{1}$ and $E_{2}$
belong to subblocks with $j=N/2$ in accordance with a theorem from
\cite{HeppLieb}. Comparing $E_{0}$ and $E_{1}$ we find $g_c=\sqrt{\omega\epsilon}$. The same condition to have
a non zero population of the bosonic mode in the ground state  was established
in \cite{scharf}. 

Comparing $E_{1}$ and $E_{2}$ near the resonance $\omega=\epsilon$
we find that a state with $c=2-N/2$ becomes the ground state 
when $g$ exceeds some value $g_{2}$ given by
\begin{equation}
g_{2}=\frac{\epsilon+\omega+\left(\omega-\epsilon\right)/
\left(2N-1\right)}{2\left(2\sqrt{1-1/ 2N }-1\right)}.
\label{eq:E2_boundary}\end{equation}
Thus,  $g<g_{2}$ together with $g>g_c$
define the regime
of the model parameters where matrix diagonalization is the only way
to study the Dicke model. The upper bound of Eq.~(\ref{eq:g_1boson})
coincides with Eq.~(\ref{eq:E2_boundary}) for $\epsilon\approx\omega$
and $N\gg 1$. 
Increasing $g$ further, we can determine the boundaries between the ground states with different values of $c$ 
( i.e. $c=3-N/2, 4-N/2, \ldots $) by numerical diagonalization, as shown in Fig.\ref{fig:gs_cooperativity} for $N=3$. A result similar to Eqs.~(\ref{eq:E1}, \ref{eq:E2}, \ref{eq:E2_boundary}) was obtained in \cite{buzek} but the descreete jumps in the cooperation number of the ground state  were interpreted as an infinite sequence of instabilities.

In the strong coupling regime, $g \gg g_c$,
the mean field approach provides a good approximation to the exact ground
state. Indeed,  introducing the expectation value of the bosonic operator $\mathcal{B}=\left\langle b\right\rangle $
and neglecting quantum fluctuations around it, the Dicke Hamiltonian
Eq.~(\ref{eq:Dicke_Model}) becomes \begin{equation}
H=\omega\left|\mathcal{B}\right|^{2}+\sum_{j}\left(\epsilon S_{j}^{z}+\frac{g}{\sqrt{N}}\left(S_{j}^{+}\mathcal{B}
+S_{j}^{-}\mathcal{B}^{*}\right)\right).\label{eq:MF_Dicke_Model}\end{equation}
Eigenstates of this Hamiltonian are product states of the N spins, and thus are manifestly not entangled. Diagonalization of the $2\times 2$ - matrices for each spin and subsequent minimization of the sum of the lowest eigenenergies over $\left|\mathcal{B}\right|$
gives the following self-consistency (mean-field) equation \begin{equation}
\omega=\frac{g^{2}}{\sqrt{4\left|\mathcal{B}\right|^{2}g^{2}/N+\epsilon^{2}}},\label{eq:MF_B}\end{equation}
which describes a quantum phase transition at a threshold value of the coupling strength $g_c$ that we have already found from the matrix diagonalization.
\begin{figure}
\centering\includegraphics[scale=0.45]{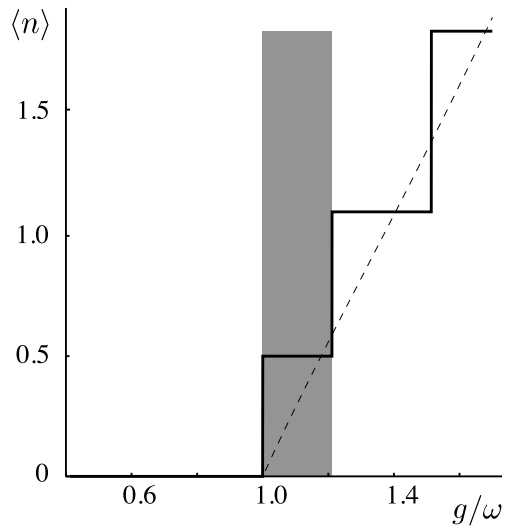}

\caption{Expectation value of $n$ calculated by matrix diagonalization of
Eqs.~(\ref{eq:diagonal_ME},\ref{eq:offdiagonal_ME}) - solid line
and the mean field result - dashed line. Plots are for $N=3$ spins and at resonance $\epsilon=\omega$. 
The grey area is the regime defined by Eq.(1) where the cooperation number of the ground state is $c=1-N/2$.\label{fig:boson_number}}

\end{figure}

The mean field approximation is satisfactory even for a system
of only a few spins. To see this, we compare  for $N=3$ spins
in Fig.\ref{fig:boson_number} 
the expectation value of $n$
taken with respect to the exact ground state
 of Eqs.~(\ref{eq:diagonal_ME},\ref{eq:offdiagonal_ME}) with 
 $\left|\mathcal{B}\right|^{2}$ given in Eq.~(\ref{eq:MF_B}) at the resonance $\omega=\epsilon$.
There we see that the largest deviation is in the region of intermediate coupling strength where the cooperation number of the exact ground state is $c=1-N/2$, grey area in Fig.\ref{fig:boson_number}. For weak coupling
$g<g_c$ the ground state coincides with the non-interacting
one and $\langle n \rangle=\left|\mathcal{B}\right|^{2}=0$. The quantum fluctuations of $n$ compared to its mean value are already small  for the coupling corresponding to the exact ground state with $c=4-N/2$,
$\left(\left\langle n^{2}\right\rangle -\left\langle n\right\rangle ^{2}\right)/\left\langle n\right\rangle ^{2}\simeq0.2$.
Thus, in the strong coupling regime the approximation of neglecting these fluctuations in Eq.~(\ref{eq:MF_Dicke_Model}) is already good for $g \gtrsim g_2$ irrespective of $N$.

The ground state can be characterized in terms of entanglement
between different parts of the spin-subsystem. In the weak coupling
regime, the ground state, being a direct product of (unexcited) spin states,
 is not entangled. In the regime of strong coupling,
where the approximate Hamiltonian Eq.~(\ref{eq:MF_Dicke_Model}) is valid, the ground
state is also a product of the  individual spin 
states which, thus, also has no entanglement of any pair of spins. The ground
state in the intermediate region Eq.~(\ref{eq:g_1boson})
has to be found by matrix diagonalization and  will be analyzed
below.

Diagonalization of the matrix Eq.~(\ref{eq:diagonal_ME},\ref{eq:offdiagonal_ME})
for $c=1-N/2$ and $g=\sqrt{\omega\epsilon}$ gives a ground state in the regime defined by Eq.~(\ref{eq:g_1boson}).  We change the mixed spin-boson representation $\left|c,l,m\right\rangle$
to a separate representation of spins and the boson $\left|\mathfrak{n},l,m\right\rangle $,
where $\mathfrak{n}$ is the bosonic occupation number. Then, tracing out the bosonic degree of freedom $\mathfrak{n}$
we obtain the reduced density matrix of the spins only, which in the representation
of $J_{z}$ eigenstates $\left|m\right\rangle $ is given by
\begin{multline}
\hat{\rho}=\frac{\epsilon}{\epsilon+\omega}\left|-\frac{N}{2}\right\rangle \left\langle -\frac{N}{2}\right|\\
+\frac{\omega}{\epsilon+\omega}\left|-\frac{N}{2}+1\right\rangle \left\langle -\frac{N}{2}+1\right|.\label{eq:rho_gs}\end{multline}
There is a finite probability to find either all spins in the completely polarized
state or in the W-state $\left|-N/2+1\right\rangle $. 

The W-state is a Bell state for $N=2$. For $N>2$ it is a bipartite entangled
state. The set of spins can be divided into two equal groups consisting of $N/2$
spins each (assuming N even). In the basis $\left|m_{1}\right\rangle \left|m_{2}\right\rangle $, $m_1$ and  $m_2$ are the eigenvalues of the operators $J_z^{1(2)}$ belonging to the first  (second) group,
the state $\left|-N/2+1\right\rangle $ is \begin{equation}
\left(\left|-\frac{N}{4}\right\rangle \left|-\frac{N}{4}+1\right\rangle 
+\left|-\frac{N}{4}+1\right\rangle \left|-\frac{N}{4}\right\rangle \right)/\sqrt{2},\label{eq:m1m2}\end{equation}
i.e. a measurement of one group projects another group onto
the definite state. Therefore in the intermediate region Eq.(1)
the ground state of the Hamiltonian Eq.~(\ref{eq:Dicke_Model}) is a bipartite entangled	
state. Note that $N$ spin W-state does not belong to the class of $N$ spin entangled states prepared by  squeezing \cite{ZollerCirac}. The squeezing parameter is not defined for  the $N=2$\; W-state and is greater than one for $N>2$\; W-states.

Let the spins have different splitting energies $\epsilon_j$ in Eq.(\ref{eq:Dicke_Model}) instead of the case $\epsilon_j\equiv\epsilon$ which we have considered in the paper. A small variation $\langle \epsilon_j \rangle-\epsilon_j\ll\epsilon_j$, $\langle \epsilon_j \rangle$ being the average over splitting energies of all spins, can be treated using  perturbation theory and will not affect our results, Eqs.(1,13) much. For instance,  the $N=2$\; W-state in Eq.(13) becomes  $\left((1-(\epsilon_1-\epsilon_2)/\sqrt{2}g)\left|\uparrow\downarrow\right\rangle+\left|\downarrow\uparrow\right\rangle\right)/\sqrt{2}$\; at the resonance $\epsilon=\omega$. The amplitudes of different components in the singlet will alter slightly while the W-state will not change qualitatively. For a more detailed analysis of the inhomogeneous Dicke model see \cite{OTLoss}.

In conclusion, we analyzed the Dicke model for a finite-size system by matrix diagonalization. We found that the ground state is a bipartite entangled state only in a narrow regime of parameters
next to the  quantum phase transition at zero temperature. The ground state
in this regime cannot be obtained from mean-field theory which
approximates the ground state as a product state. Also perturbation theory is not valid in this regime. For an infinite system width of the corresponding parametric region vanishes as $1/N$ if system is close the resonance.

We acknowledge financial support
from the Swiss NF, NCCR Nanoscience Basel, and JST ICORP.


\begin{thebibliography}{10}

\bibitem{Gardiner} C. W. Gardiner and P. Zoller, \emph{Quantum Noise}, Springer, 2004.

\bibitem{esslinger}F. Brennecke, T. Donner, S. Ritter,
T. Bourdel, M. K\"{o}hl, and T. Esslinger, Nature \textbf{450},
268 (2007).

\bibitem{kasprzak}J. Kasprzak, M. Richard, S. Kundermann, A. Baas,
P. Jeambrun, J. M. J. Keeling, F. M. Marchetti, M. H. Szyma\`{n}ska, R.
Andr\`{e}, J. L. Staehli, V. Savona, P. B. Littlewood, B. Deveaud, and
Le Si Dang, Nature \textbf{443}, 409 (2006).

\bibitem{wallraff}A. Wallraff, D. I. Schuster, A. Blais, L. Frunzio,
R.- S. Huang, J. Majer, S. Kumar, S. M. Girvin and R. J. Schoelkopf,
Nature \textbf{431}, 162 (2004) .

\bibitem{Badolato1}K. Hennessy, A. Badolato, M. Winger, D. Gerace,
M. Atat\"{u}re, S. Gulde, S. F\"{a}lt, E. L. Hu, and A. \.{I}mamo\u{g}lu, Nature \textbf{445},
896 (2007).

\bibitem{AwschalomQED} A. \.{I}mamo\u{g}lu, D. D. Awschalom, G. Burkard, D. P. DiVincenzo, D. Loss, M. Sherwin, and A. Small, Phys. Rev. Lett. \textbf{83}, 4204 (1999).

\bibitem{Dicke}R. H. Dicke, Phys. Rev. \textbf{93}, 99 (1954).

\bibitem{TavisCummings}M. Tavis and F. W. Cummings, Phys. Rev. \textbf{170}, 379 (1968).

\bibitem{scharf}G. Scharf, Helv. Phys. Acta \textbf{43}, 806(1970).

\bibitem{HeppLieb}K. Hepp and E. H. Lieb, Ann. Phys. (N.Y.) \textbf{76}, 360 (1973).

\bibitem{eastham}P. R. Eastham and P. B. Littlewood, Phys. Rev. B \textbf{64}, 235101 (2001).

\bibitem{MeanField}R. Bonifacio and G. Preparata, Phys. Rev. A \textbf{2}, 336 (1970).
\bibitem{Pratt} J. S. Pratt, Phys. Rev. B \textbf{73}, 184413 (2006).

\bibitem{spin_division}Independent probe of different spins is possible
as they are spatially separated. For instance, size of the exciton-polariton
condensates in \cite{kasprzak} is few $\mu m$; cloud of cold atoms
in \cite{wallraff} has cigar shape of length that can be 100 $\mu m$. 

\bibitem{SW_details} The Hamiltonian Eq.(\ref{eq:H_SW}) was obtained using  perturbation theory for $\epsilon \approx \omega \gg g^2/(\epsilon-\omega)$. Expressing XY term from the total spin conseravtion law the Hamiltonian Eq.(\ref{eq:H_SW}) is
$ H=\omega b^{\dagger}b+\left(\epsilon+\frac{g^{2}(2 b^{\dagger}b+1)}{N\left(\epsilon-\omega\right)}\right)\sum_{j}S_{j}^{z} 
-\frac{g^{2}}{N\left(\epsilon-\omega\right)}\left(\sum_{j}S_{j}^{z}\right)^2 $
Its ground state is all spin polarized up within the applicability limits of perturbation theory. The same ground state is predicted by mean-field theory in this regime.

\bibitem{mircea}M. Trif, V. N. Golovach, and D. Loss,
Phys. Rev. B \textbf{77}, 045434 (2008).


\bibitem{buzek} Vladimir Buzek, Miguel Orszag, and Marian Rosko, Phys. Rev. Lett. {\bf 94}, 163601 (2005).

\bibitem{ZollerCirac} A. S{\o}rensen, L.-M. Duan, J. I. Cirac, and P. Zoller,  Nature  {\bf 409}, 63 (2001)

\bibitem{OTLoss} O. Tsyplyatyev and D. Loss, to appear in PRA.

\end{thebibliography}
\end{document}